\documentclass[12pt,italian,english]{article}
\usepackage{lmodern}

\usepackage[T1]{fontenc}
\usepackage[latin9]{inputenc}
\usepackage{geometry}
\usepackage{color}
\usepackage{float}
\usepackage{amsthm}
\usepackage{amsmath}
\usepackage{amssymb}
\usepackage{fixltx2e}
\usepackage{graphicx}
\usepackage{setspace}
\usepackage{esint}
\makeatletter

\usepackage{slashed}

\usepackage{amsthm}
\usepackage{graphicx}
\usepackage{amsmath}
\usepackage{amssymb}
\usepackage{amsfonts}
\usepackage[british]{babel}
\usepackage{slashed}
\usepackage{afterpage}
\usepackage{cite}
\usepackage{color}
\usepackage{pdfsync}
\def\appendix#1{\addtocounter{section}{1}\setcounter{equation}{0}
\renewcommand{\thesection}{\Alph{section}}
\section*{
\thesection\protect\indent \parbox[t]{11.715cm} {#1}}
\addcontentsline{toc}{section}{Appendix\thesection\ \ \ #1} }

\font\mybb=msbm10 at 12pt
\def\bb#1{\hbox{\mybb#1}}

\def\be{\begin{equation}}
\def\ee{\end{equation}}
\def\bea{\begin{eqnarray}}
\def\eea{\end{eqnarray}}

\makeatother

\usepackage{babel}
\begin{document}
 \begin{titlepage}~\\
~\\
~

\begin{center}
\textbf{\Large Spectral Action and Gravitational effects }\\
\textbf{\Large at the Planck scale}
\par\end{center}{\Large \par}

\begin{center}
{Agostino Devastato$^{1,2}$} \\[6mm] $^{1}$\textit{Dipartimento
di Fisica, Universit� di Napoli }\textsl{Federico II} \\[4mm] $^{2}$\textit{INFN,
Sezione di Napoli}\\
 \textit{Monte S.~Angelo, Via Cintia, 80126 Napoli, Italy}\\
~
\par\end{center}

\begin{center}
\texttt{\small agostino.devastato@na.infn.it}\texttt{ }
\par\end{center}

\vskip 2 cm
\begin{abstract}
We discuss the possibility to extend the spectral action up to energy
close to the Planck scale, taking also into account the gravitational
effects given by graviton exchange. Including this contribution in
the theory, the coupling constant unifi{}cation is not compromised,
but is shifted to the Planck scale rendering all gauge couplings asymptotically
free. In the scheme of noncommutative geometry, the gravitational
effects change the main standard model coupling constants, leading
to a restriction of the free parameters of the theory compatible with
the Higgs and top mass prediction. We also discuss consequences for
the neutrino mass and the see-saw mechanism.
\end{abstract}
 \end{titlepage}

\section{Introduction}

Noncommutative geometry~\cite{Connesbook, Landi, Ticos, ConnesMarcolli}
allows to handle a large variety of geometrical frameworks from a
totally algebraic point of view. In particular it is very useful in
the derivation of models in high energy physics, such as the Yang-Mills
gauge theories~\cite{ConnesLott, Schucker, AC2M2, Walterreview,Jureit}.
In the current state the noncommutative geometry structure of gauge
theories is understood to be an ``almost commutative'' geometry,
i.e. the prduct of continuous geometry, representing space-time, times
an internal algebra of finite dimensional matrix. In this geometric
framework the spectral action principle~\cite{spectralaction} enables
the retrieval of the full standard model of high energy physics, including
the Higgs field: the standard model is put on the same footing as
geometrical general relativity making it a possible unification with
gravity. In fact the application of noncommutative geometry to gauge
theories of strong and electroweak forces is a very original way to
fully geometrize the interaction of elementary particles. Furthermore
it has been shown~\cite{ccforgotafield} that it is possible to extend
the standard model by including an additional singlet scalar field
that stabilizes the running coupling constants of the Higgs field.
This singlet scalar field is closely related to the right-handed Majorana
neutrinos, conferring them mass, and leading to the prediction of
the seesaw mechanism which explains the large difference between the
masses of neutrinos and those of the other fermions. A recent model~\cite{DLM}
shows the possibility of a further extension, going one step higher
in the construction of the noncommutative manifold, in a sort of noncommutative
geometry grand unification: here it is pointed out that there could
be a ``next level'' in noncommutative geometry, intertwined with
the Riemannian and spin structure of spacetime, where the singlet-scalar
field rises. Accordingly it naturally appears at high scale, near
to the Planck scale.

A possible framework for describing interactions at energies and momenta
below the Planck scale is given in~\cite{RobinsonWilczek,Shaposhnikov:2009pv}.
Therefore in this paper we check the possibility to extend the unification
scale up to the Planck scale $M_{P}\equiv\sqrt{\hbar c/G_{N}}\simeq10^{19}\mbox{GeV}$,
including not neglible gravitational effects. Moreover for a theory
dealing with the unifi{}cation of gauge theory and gravity a more
natural scale is the Planck scale. The usual strategy is to use the
spectral action as an effective action at a fixed scale, of the order
of the unification scale, and to impose the additional relations between
the independent parameters of the standard model. Then, using the
RG equations, one can let these parameters run to their value at low
scales and evaluate the Higgs, the top and neutrino masses. The question
here is: what is the predictive power of this extended model with
exchange of gravitons at the Planck scale? We want to see how the
gravitational effects change the main running coupling constants and
if they lead to a restriction on the free parameters of the theory
still compatible with the Higgs, top and neutrino mass predictions.

In~\cite{Estrada:2012te} Marcolli and Estrada carried out a similar
analysis within the asymptotic safety scenario with Gaussian matter
fixed point; differently from this paper, they have not considered
the effect of the the scalar field $\sigma$ introduced in~\cite{ccforgotafield}
, which is necessary in order to reproduce the seesaw mechanism and
to have the Higgs mass with its correct value.

The present paper is organized as follows. In section 2 some ingredients
and the main results of the spectral action principle are shown: the
derivation of the full standard model bosonic action plus the singlet
scalar field and gravity. In sect. 3 the gravitational contributions
to the three gauge couplings, not neglible at the Planck scale, are
presented. In sect. 4 is shown how the gravitational effects change
the RG equations of the Yukawa and autointeraction Higgs couplings
leading to a restriction of the free parameters of the theory compatible
with the Higgs and top mass. The fi{}nal section contains conclusions
and some comments.

\section{The spectral action}

We recall the main features of the spectral action, referring to the
original works~\cite{Connesbook,spectralaction} for the full treatment.
Those familiar with this calculation can skip to the next section.

The basic ingredients of noncommutative geometry are an algebra $\mathcal{A}$,
which involves the topology of space-time and its noncommutative generalization;
an Hilbert space $\mathcal{H}$ on which the algebra acts, containing
the fermionic degrees of freedom; and a generalized Dirac operator
$D$ which encodes the metric structure of the space. These three
objects form the so called spectral triple. The triple is said to
be even if there is an operator $\Gamma$ on $\mathcal{H}$ such that
$\Gamma=\Gamma^{*},\,\Gamma^{2}=1$ and
\begin{eqnarray}
\Gamma D+D\Gamma=0 & \,; & \,\Gamma a-a\Gamma=0\,,\,\,\forall a\in\mathcal{A}\,.
\end{eqnarray}
 A spectral triple, enlarged with an anti-unitary operator $J$ on
$\mathcal{H}$ that obey  1) $J^2 = \pm \bb I$; 2) $JD= \pm DJ$; 3) $J\Gamma~=\pm\Gamma J$
(with choice of signs dictated by the $KO$-dimension of the spectral
triple), is said to be real. A real even spectral triple defines a
gauge theory, with the gauge fields arising as the inner fluctuations
of the Dirac operator: \foreignlanguage{italian}{
\begin{equation}
D_{A}=D+A+JAJ
\end{equation}
}where $A$ is the one form connection given by the commutator of
the Dirac operator $D$ and the elements of the algebra, $A=\sum_{i}a_{i}\left[D,b_{i}\right];$
the Dirac operator is the product of a continuous part representing
space-time, times an internal part of finite dimensional matrix:
\begin{equation}
D=\slashed{\partial}_{\omega}\otimes\mathbb{I}_{F}+\gamma^{5}\otimes D_{F}\label{eq:D operatore MS-1-1-1}
\end{equation}
where $\slashed{\partial}_{\omega}\equiv\gamma^{\mu}\left(\partial_{\mu}+\omega_{\mu}\right)$
and
\begin{equation}
D_{F}=\left(\begin{array}{cccc}
0 & \mathcal{M} & \mathcal{M}_{R} & 0\\
\mathcal{M}^{\dagger} & 0 & 0 & 0\\
\mathcal{M}_{R}^{\dagger} & 0 & 0 & \mathcal{M}^{*}\\
0 & 0 & \mathcal{M}^{T} & 0
\end{array}\right)\,,\,\mbox{with\,}\,\begin{array}{c}
\mathcal{M}=\left(\begin{array}{cc}
M_{\nu} & 0\\
0 & M_{l}
\end{array}\right)\\
\mathcal{M}_{R}=\left(\begin{array}{cc}
M_{R} & 0\\
0 & 0
\end{array}\right)
\end{array}\label{eq:D_F Modello Standard-1-1-1}
\end{equation}
The matrices $\mathcal{M}$ and $\mathcal{M}_{R}$, via $M_{l},\, M_{\nu},\,\mbox{and}\, M_{R}$,
contain respectively Dirac and Majorana masses, or better Yukawa couplings
of leptons, Dirac and Majorana neutrinos. 

By the Dirac operator $D_{A}$ we deduce the full bosonic action of
high energy physics coupled to gravity~\cite[Sect.~4.1]{AC2M2} through
the regularization of its eigenvalues,
\begin{equation}
S_{B}[A]\equiv\mbox{Tr}f\left(\frac{D_{A}^{2}}{\Lambda^{2}}\right)\,
\end{equation}
where $f$ is a smooth cut-off function and $\Lambda$ is the cut-off
scale of the order of the unification scale. The parameter $\Lambda$
is used to obtain an asymptotic series for the spectral action via
the heath kernel expansion; the physically relevant terms appear with
a positive power of $\Lambda$ as coefficient.  One could show that
this bosonic action is derivable from its fermionic counterpart via
the renormalization flow in the presence of anomalies~\cite{AndrianovLizzi,AndrianovKurkovLizzi,Kurkov:2012dn}.
The fermionic action is given by
\begin{equation}
S_{F}=\overline{J\psi}\left(D+A+JAJ\right)\psi\,.
\end{equation}

Now let us see the form of the action starting from the formula for
a second-order elliptic differential operator $D_{A}^{2}$ of the
form 
\begin{equation}
D_{A}^{2}=-(g^{\mu\nu}\partial_{\mu}\partial_{\nu}+K^{\mu}\partial_{\mu}+L)\,.
\end{equation}
This operator can be written using a connection $\nabla_{\mu}$ so
that 
\begin{equation}
D_{A}^{2}=-\left(g^{\mu\nu}\nabla_{\mu}\nabla_{\nu}+E\right)\label{eq: Altra forma di D^2-1}
\end{equation}
Explicitly, $\nabla_{\mu}=\nabla_{\mu}^{[R]}+\omega{}_{\mu}$ contains
both Riemann $\nabla_{\mu}^{[R]}$ and ``gauge'' $\omega$ parts,
with 
\begin{equation}
\omega{}_{\mu}=\frac{1}{2}g_{\mu\nu}\left(K^{\nu}+g^{\rho\sigma}\Gamma_{\rho\sigma}^{\nu}\right).\label{eq: omega mu-1}
\end{equation}
Using this $\omega{}_{\mu}$ and $L$ we find $E$ and compute the
curvature $\Omega_{\mu\nu}$ of $\nabla$: 
\begin{eqnarray}
E & \equiv & L-g^{\mu\nu}\partial_{\nu}(\omega{}_{\mu})-g^{\mu\nu}\omega{}_{\mu}\omega{}_{\nu}+g^{\mu\nu}\omega{}_{\rho}\Gamma_{\mu\nu}^{\rho}\,;\nonumber \\
\Omega_{\mu\nu} & \equiv & \partial_{\mu}(\omega{}_{\nu})-\partial_{\nu}(\omega{}_{\mu})-\left[\omega{}_{\mu},\,\omega_{\nu}\right]\,.\label{eq:def Omega-1}
\end{eqnarray}

The spectral action has an heath kernel expansion in a power series
in terms of $\Lambda^{-1}$ as
\begin{equation}
\mbox{Tr}f\left(\frac{D_{A}^{2}}{\Lambda^{2}}\right)=2\Lambda^{4}f_{0}a_{0}(D_{A}^{2})+2\Lambda^{2}f_{2}a_{2}(D_{A}^{2})+f_{4}a_{4}(D_{A}^{2})+O(\Lambda^{-2})\,,\label{eq:Traccia-1}
\end{equation}
where the $f_{k}$ are momenta of the function $f$, 
\begin{eqnarray}
f_{0} & = & \int_{0}^{\infty}uf(u)du\,,\,\, f_{2}=\int_{0}^{\infty}f(u)du\,,\,\, f_{2n+4}=(-)^{n}\partial_{u}^{n}f(u)
\end{eqnarray}
and the coefficients $a_{n}(x,P)$ are called the Seeley-DeWitt coefficients~\cite{Gilkey, Vassilevich}.
They are equal to zero for $n$ odd and the first three even coefficients
are given by 
\begin{eqnarray}
a_{0}(x,P) & = & (4\pi)^{-m/2}\mbox{Tr}(Id)\nonumber \\
a_{2}(x,P) & = & (4\pi)^{-m/2}\mbox{Tr}(-R/6\, Id+E)\nonumber \\
a_{4}(x,P) & = & (4\pi)^{-m/2}\mbox{Tr}(-12R_{;\mu}^{\mu}+5R^{2}-12R_{\mu\nu}R^{\mu\nu}-60RE+180E^{2}\nonumber \\
 &  & \,\,\,\,\,\,\,\,\,\,\,\,\,\,\,\,\:+R_{\mu\nu\rho\sigma}R^{\mu\nu\rho\sigma}+60E_{;\mu}^{\mu}+30\Omega_{\mu\nu}\Omega^{\mu\nu})\label{eq:SeelydeWitt}
\end{eqnarray}

\section{Higgs-singlet scalar potential and Gravity}

By inserting relations for the Seeley-DeWitt coefficients (\ref{eq:SeelydeWitt})
into (\ref{eq:Traccia-1}) we obtain the standard model action plus
a new singlet scalar field coupled to gravity\cite[Eq.(5.49)]{CCpart1}:
\begin{flalign}
S_{B} & =\frac{24}{\pi^{2}}f_{4}\Lambda^{4}\int d^{4}x\sqrt{g}-\frac{2}{\pi^{2}}f_{2}\Lambda^{2}\int d^{4}x\sqrt{g}\left[R+\frac{1}{2}a\overline{H}H+\frac{1}{4}c\sigma^{2}\right]+\\
 & +\frac{1}{2\pi^{2}}f_{0}\int d^{4}x\sqrt{g}[\frac{1}{30}\left(-18C_{\mu\nu\rho\sigma}^{2}+11R^{*}R^{*}\right)+\frac{5}{3}g_{1}^{2}B_{\mu\nu}^{2}+g_{2}^{2}\mathbf{W}_{\mu\nu}^{2}+g_{3}^{2}\mathbf{V}_{\mu\nu}^{2}\nonumber \\
 & +\frac{1}{6}aR\overline{H}H+b\left(\overline{H}H\right)^{2}+a(\nabla_{\mu}H)^{2}+2e\overline{H}H\sigma^{2}+\frac{1}{2}d\sigma^{4}+\frac{1}{12}cR\sigma^{2}+\frac{1}{2}c\left(\partial_{\mu}\sigma\right)^{2}]\nonumber 
\end{flalign}
where $B_{\mu\nu}$, $\mathbf{W}_{\mu\nu}$ and $\mathbf{V}_{\mu\nu}$
are respectivly the field strenght associated with the gauge groups
$U(1),\, SU(2)$ and $SU(3)$; $H$ is identified with the Higgs field
and $\sigma$ is a singlet-scalar field. This field is related to
the neutrino Majorana mass that allows to reproduce a seesaw mechanism
of I type as described in~\cite{AC2M2}. Furthermore, this $\sigma$
field lowers the standard model Higgs mass to its experimental value.
The three momenta $f_{0},\, f_{2}\mbox{ and }f_{4}$ can be used to
specify the initial conditions of the gauge couplings, the Newton
constant and the cosmological constant. The coefficients $a,b,c,d\mbox{ and }e$
are related to the fermionic Yukawa couplings and Majorana mass matrix
and will be written in the crude approximation where the Yukawa couplings
of the top quark $y_{top}$ and the neutrino (both Majorana $y_{\nu_{R}}$and
Dirac $y_{\nu}$) are dominant; in addition, we introduce the dimensionless
constant $\rho$ defined by the ratio between the Dirac Yukawa couplings
$y_{\nu}=\rho y_{top}$: {\small 
\begin{align}
a & =\mbox{tr}\left[y_{\nu}^{*}y_{\nu}+y_{e}^{*}y_{e}+3\left(y_{top}^{*}y_{top}+y_{d}^{*}y_{d}\right)\right]\simeq(3+\rho^{2})y_{top}^{2}\nonumber \\
b & =\mbox{tr}\left[\left(y_{\nu}^{*}y_{\nu}\right)^{2}+\left(y_{e}^{*}y_{e}\right)^{2}+3\left(y_{top}^{*}y_{top}+y_{d}^{*}y_{d}\right)^{2}\right]\simeq(3+\rho^{4})y_{top}^{4}\nonumber \\
c & =\mbox{tr}\left[y_{\nu_{R}}^{*}y_{\nu_{R}}\right]\simeq y_{\nu_{R}}^{2}\nonumber \\
d & =\mbox{tr}\left[\left(y_{\nu_{R}}^{*}y_{\nu_{R}}\right)^{2}\right]\simeq y_{\nu_{R}}^{4}\nonumber \\
e & =\mbox{tr}\left[y_{\nu}^{*}y_{\nu}y_{\nu_{R}}^{*}y_{\nu_{R}}\right]\simeq\rho^{2}y_{top}^{2}y_{\nu_{R}}^{2}
\end{align}
}Furthemore it is more transparent to work with the rescaled fields
\begin{equation}
H\rightarrow\left(\sqrt{\frac{2}{3+\rho^{2}}}g\right)\frac{H}{y_{top}}\mbox{ \,\,\,\,\,;\,\,\,}\,\sigma\rightarrow(2g)\frac{\sigma}{y_{\nu_{R}}}
\end{equation}
 (where $g$ is the gauge coupling to the unification scale) so that
the spectral action for scalar fields and gravity reduces to {\small 
\begin{multline}
S_{B}=\frac{24}{\pi^{2}}f_{4}\Lambda^{4}\int d^{4}x\sqrt{g}-\frac{2}{\pi^{2}}f_{2}\Lambda^{2}\int d^{4}x\sqrt{g}\left[R+g^{2}H^{2}+g^{2}\sigma^{2}\right]+\frac{1}{2\pi^{2}}f_{0}\int d^{4}x\sqrt{g}\left[\left(\frac{4}{3+\rho^{2}}\right)g^{4}H^{4}\right.\\
\left.+2(\nabla_{\mu}H)^{2}+8g^{4}\frac{2\rho^{2}}{3+\rho^{2}}H^{2}\sigma^{2}+8g^{4}\sigma^{4}+2g^{2}\left(\partial_{\mu}\sigma\right)^{2}+\frac{1}{3}g^{2}R\left(H^{2}+\sigma^{2}\right)\right].
\end{multline}
}\textcolor{black}{In the action above we have neglected the additional
gravitational term given by the Weyl curvature. This term is subdominant
to the Einstein-Hilbert term at unification scale \cite{Marcolli:2009in}.
It could be shown \cite{AC2M2} that the running of this term changes
by at most an order of magnitude at lower scales, so we can assume
that it remains subdominant and neglect it in first approximation.
Moreover we are neglecting the quadratic term in $R$.}\\
By setting the coefficient $f_{0}$ to be $\frac{1}{2\pi^{2}}f_{0}=\frac{1}{4g^{2}}$
one obtains the normalization of the gauge fields kinetic terms so
that the Higgs-singlet potential plus gravity reduces to
\begin{align}
V & =\frac{1}{4}\left(\lambda_{H}H^{4}+\lambda_{\sigma}\sigma^{4}+2\lambda_{H\sigma}H^{2}\sigma^{2}\right)-\frac{2g^{2}}{\pi^{2}}f_{2}\Lambda^{2}\left(H^{2}+\sigma^{2}\right)\nonumber \\
 & +\frac{1}{12}R\left(H^{2}+\sigma^{2}\right)-\frac{2}{\pi^{2}}f_{2}\Lambda^{2}R+\frac{24}{\pi^{2}}f_{4}\Lambda^{4}
\end{align}
where $\lambda_{H},\,\lambda_{\sigma},\,\lambda_{H\sigma}$ are defined
in terms of $g,$ that is the value of the three coupling constants
at the unification scale,
\begin{eqnarray}
\lambda_{H} & \equiv & \frac{\rho^{4}+3}{\left(3+\rho^{2}\right)^{2}}4g^{2}\,;\,\,\,\lambda_{H\sigma}\equiv\frac{2\rho^{2}}{\rho^{2}+3}4g^{2}\,;\,\,\,\lambda_{\sigma}\equiv8g^{2}\,.\label{eq:lambda h unificazione}
\end{eqnarray}
The usual strategy, at this point, is to use the spectral action as
an effective action at a fixed scale, of the order of the GUT scale$\simeq10^{17}GeV$,
and to impose the additional relations (\ref{eq:lambda h unificazione})
between the independent parameters of the standard model as a boundary
condition at that scale. Differently, in the following, we shift the
unification scale to the Planck scale $M_{P}$. Hence, we want to
study the framework in which general relativity is quantized for small
fl{}uctuations around a fl{}at space-time and the Planck scale becomes
the real unification scale of all physical interactions. In this extension
of the spectral action to higher energy scales, we will include the
contribution of graviton exchange in the running coupling constants.
Of course, these contributions will not be significant for low energies
and they will be only important near the Planck scale. By using these
new RG equations we can let the standard model parameters run to their
value at low scale and test the predictive power of the model: we
will obtain a constrain of the free parameters of the theory still
compatible with the Higgs and top mass prediction.

\section{Gravitational correction to running of Gauge couplings}

\begin{figure}
\centering{}\includegraphics[width=4cm,height=4cm]{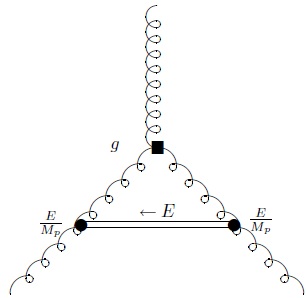}\caption{\label{fig:A-typical-Feynman}\textit{A typical Feynman diagram at
one-loop for a gravitational process contributing to the gauge coupling
renormalization. Double lines represent gravitons. Curly lines represent
gluons. The three-gluon vertex $\blacksquare$ is proportional to
$g_{i}$, while the gluon-graviton vertex $\bullet$ is proportional
to$E/M_{P}$.}}
\end{figure}

\begin{figure}
\centering{}\includegraphics[width=9cm,height=5cm]{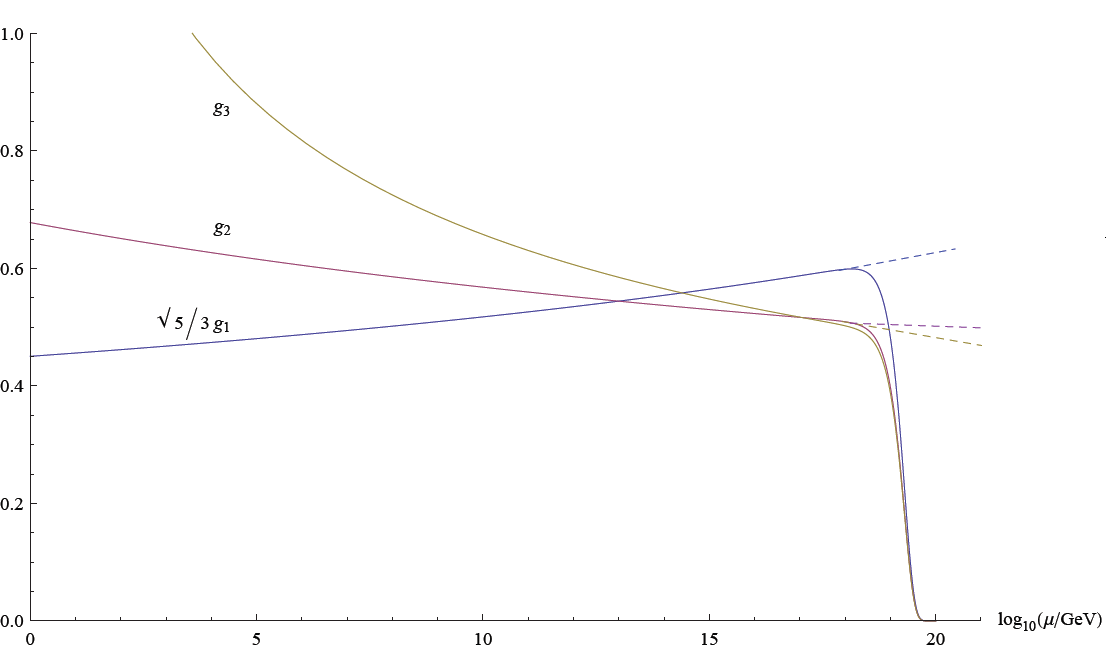}\caption{\label{fig:gauge couplings}\textit{Including gravity at one-loop,
the couplings remain unifi{}ed near $10^{17}$ GeV, but evolve rapidly
to zero at high $E$.}}
\end{figure}

A possible framework for describing interactions at energies and momenta
below the Planck scale is given in~\cite{RobinsonWilczek}. The dynamics
for a non-Abelian gauge fi{}eld coupled to gravity is given by the
action,
\begin{equation}
\int d^{4}x\sqrt{g}\left[\frac{1}{k_{pl}^{2}}R-\frac{1}{4g^{2}}\left(\frac{5}{3}g_{1}^{2}B_{\mu\nu}^{2}+g_{2}^{2}\mathbf{W}_{\mu\nu}^{2}+g_{3}^{2}\mathbf{V}_{\mu\nu}^{2}\right)\right]\,.
\end{equation}
where we have used the momentum $f_{2}$ to specify the initial conditions
of the Planck constant, $\frac{2}{\pi^{2}}f_{2}\Lambda^{2}\equiv\frac{1}{k_{pl}^{2}}\equiv M_{P}^{2}/16\pi.$
The form of the gravitational correction can be determined on general
grounds, involving in the one-loop Feynman diagrams of interest a
gluon vertex dressed by exchange of gravitons (See Fig.\ref{fig:A-typical-Feynman}).
 Since the gauge boson vertex has strength $g_{i}$ and gravitons
couple to energy-momentum with a dimensional coupling $\propto1/M_{P}$,
dimensional analysis implies that the running of couplings in four
dimensions will be governed by a Callan-Symanzik $\beta$ function
of the form~\cite[Eq.~19]{RobinsonWilczek} 
\begin{equation}
\beta(g_{i},E)=\frac{b_{i}}{16\pi^{2}}g_{i}^{3}+a_{g}\frac{E^{2}}{M_{P}^{2}}g_{i}\,\,,\,\mbox{con }b_{i}=(\frac{41}{6},-\frac{19}{6},-7)\label{eq:beta gravitons}
\end{equation}
where the first term represents the usual standard model contribution
and the second one includes the gravitational correction. Initial
values of $g_{i}$ are set with the experimental values at $M_{Z}\simeq91\,\mbox{GeV}$:
$g_{1}(M_{Z})=0.3575,\, g_{2}(M_{Z})=0.6514,\, g_{3}(M_{Z})=1.221$.
The numerical value of $a_{g}$, also called \textit{anomalous dimension},
is determined by a detailed calculation described in~\cite{RobinsonWilczek}
leading to $a_{g}=-3/\pi$ which we can rewrite $a_{g}=-\frac{3}{16\pi^{2}}k_{pl}^{2}M_{P}^{2}$.
The negative sign of this coefficient means that the gravitational
correction works in the direction of asymptotic freedom: it forces
the couplings to decrease at large energy, as it is shown in fig (\ref{fig:gauge couplings}).
At one-loop order, when gravity is ignored, the three gauge couplings
evolve as the inverse logarithm of $E$ (dashed curves); when gravity
is included, see the solid lines, the couplings evolve rapidly towards
weaker coupling at high $E$. Of course, its effect only becomes quantitatively
important when the energy approaches the Planck scale, and graviton
exchanges are no longer negligible. We finally note that the three
gauge coupling constants approximately assume the same value, about
zero, from $E\geq3\times10^{19}\mbox{GeV}$. Near the Planck scale
$E\simeq10^{19}\mbox{GeV}$ the three gauge couplings are not exactly
equal: we have $g_{1}(\Lambda)=0.372$, $g_{3}(\Lambda)=0.386$ and
$g_{2}(\Lambda)=0.396$. 

The unifi{}cation of the gauge coupling constants, at the Planck scale,
has been also considered in several frameworks ~\cite{Maiani,AndrianovEspriu}
with the request of new fermions, in a different perspective from
us.

\section{Renormalization group equations with gravitational corrections}

The running of the Higgs mass with the presence of a scalar field
has been studied in~\cite{ccforgotafield}. However the RG equations
for the matter sector have to be adapted via the addition of the anomalous
dimensions of the running parameters, that take into account the contribution
of gravity~\cite{Shaposhnikov:2009pv},
\begin{equation}
\frac{dx_{i}}{dt}=\beta_{x_{i}}^{SM}+\beta_{x_{i}}^{grav}
\end{equation}
where $x_{i}$ are the running parameters, $\beta_{x_{i}}^{SM}$ is
the Standard Model beta function for $x_{i}$ and $\beta_{x_{i}}^{grav}$
is the gravitational correction. The latter is of the general form,
\begin{equation}
\beta_{x_{i}}^{grav}=a_{x_{i}}\frac{E^{2}}{8\pi M_{P}^{2}}x_{i}(t)
\end{equation}
In our analysis we use an estimate of the anomalous dimensions as
suggested in~\cite{Shaposhnikov:2009pv}: $a_{x_{j}}$ are fixed to
1 for the Yukawa couplings and to $3.1$ for the autointeraction couplings
of the scalar fields. 

For the analysis of the renormalization group flow we shall expand
the approach presented in~\cite{Walterreview,Jureit} with the presence
of gravitational contributions. Let $M_{R}$ be the Majorana mass
for the right-handed tau-neutrino. By the Appequist-Carazzone decoupling
theorem~\cite{AppelquistCarazzone}, we can distinguish two different
energy domains: $E>M_{R}$ and $E<M_{R}$. 

For high energies $E>M_{R}$, the renormalization group equations
are given by~\cite[Eq.15]{Antusch}, \cite[Eq. B.4]{Machacek2} and
\cite[Eq. B.3]{Machacek3}, adapted via the addition of the gravitational
contributions described above
\begin{eqnarray}
\frac{dy_{top}}{dt} & = & \frac{y_{top}}{16\pi^{2}}\left(\frac{9}{2}y_{top}^{2}+y_{\nu}^{2}-\frac{17}{12}g_{1}^{2}-\frac{9}{4}g_{2}^{2}-8g_{3}^{2}\right)-a_{y_{top}}\frac{E^{2}}{8\pi M_{P}^{2}}y_{top}\nonumber \\
\frac{dy_{\nu}}{dt} & = & \frac{y_{\nu}}{16\pi^{2}}\left(3y_{top}^{2}+\frac{5}{2}y_{\nu}^{2}-\frac{3}{4}g_{1}^{2}-\frac{9}{4}g_{2}^{2}\right)-a_{y_{\nu}}\frac{E^{2}}{8\pi M_{P}^{2}}y_{\nu}\nonumber \\
\frac{d\lambda_{H}}{dt} & = & \frac{1}{16\pi^{2}}\left(24\lambda_{H}^{2}-\left(3g_{1}^{2}+9g_{2}^{2}\right)\lambda_{H}+2\lambda_{H\sigma}^{2}+\frac{6}{16}\left(g_{1}^{4}+2g_{1}^{2}g_{2}^{2}+3g_{2}^{4}\right)\right.\nonumber \\
\frac{d\lambda_{H\sigma}}{dt} & = & \frac{1}{16\pi^{2}}\left(6y_{top}^{2}+2y_{\nu}^{2}-\frac{3}{2}g_{1}^{2}-\frac{9}{2}g_{2}^{2}+12\lambda_{H}+6\lambda_{\sigma}+8\lambda_{H\sigma}\right)\lambda_{H\sigma}+a_{\lambda_{H\sigma}}\frac{E^{2}}{8\pi M_{P}^{2}}\lambda_{H\sigma}\nonumber \\
\frac{d\lambda_{\sigma}}{dt} & = & \frac{1}{16\pi^{2}}\left(8\lambda_{H\sigma}^{2}+18\lambda_{\sigma}^{2}\right)+a_{\lambda_{\sigma}}\frac{E^{2}}{8\pi M_{P}^{2}}\lambda_{\sigma}\label{eq:RG equations inizio}
\end{eqnarray}
with $E=E(t)=m_{Z}e^{t}.$ Below the threshold $E=M_{R}$, the tau-neutrino
Yukawa coupling is replaced by an effective coupling~\cite[Eq.14]{Antusch}
\begin{equation}
\kappa=2\frac{y_{\nu}^{2}}{M_{R}}\,,\label{eq: constant k}
\end{equation}
which gives an effective mass $m_{l}=\frac{1}{4}\kappa v_{0}^{2}$
to the light tau-neutrino. In the range $0<E<M_{R}$ the renormalization
group equations for $\lambda_{\sigma}$ and $\lambda_{H\sigma}$ are
the same, whereas the ones for $y_{top}$,$y_{\nu},$ and $\lambda_{H}$
are replaced by
\begin{eqnarray}
\frac{dy_{top}}{dt} & = & \frac{1}{16\pi^{2}}\left(\frac{9}{2}y_{top}^{2}-\frac{17}{12}g_{1}^{2}-\frac{9}{4}g_{2}^{2}-8g_{3}^{2}\right)-\frac{a_{y}E^{2}}{8\pi M_{P}^{2}}y_{top}\nonumber \\
\frac{d\kappa}{dt} & = & \frac{1}{16\pi^{2}}\left(6y_{top}^{2}+\frac{1}{36}\lambda_{H}-3g_{2}^{2}\right)\kappa-\frac{a_{y}E^{2}}{8\pi M_{P}^{2}}\kappa\nonumber \\
\frac{d\lambda_{H}}{dt} & = & \frac{1}{16\pi^{2}}\left(24\lambda_{H}^{2}-\left(3g_{1}^{2}+9g_{2}^{2}\right)\lambda_{H}+2\lambda_{H\sigma}^{2}+\frac{6}{16}\left(g_{1}^{4}+2g_{1}^{2}g_{2}^{2}+3g_{2}^{4}\right)+\right.\nonumber \\
 &  & \left.+12y_{top}^{2}\lambda-3y_{top}^{4}\right)+\frac{a_{\lambda_{H}}E^{2}}{8\pi M_{P}^{2}}\lambda_{H}\label{eq:RG equation fine}
\end{eqnarray}
 The numerical solutions to the coupled differential equations (\ref{eq:RG equations inizio})
to (\ref{eq:RG equation fine}) depend on three input parameters:
(1) the unification scale $\Lambda$; (2) the Majorana mass $M_{R}$
which produces the threshold in the renormalization group flow; (3)
the ratio $\rho$ between the Dirac Yukawa couplings of the top quark
and neutrino. 

The scale $\Lambda$, usually taken at the unification $\Lambda_{12}=10^{13}GeV$
or $\Lambda_{23}=10^{17}GeV$ i.e. the two extreme point in which
$g_{1}=g_{2}$ and $g_{2}=g_{3},$ is now shifted to the Planck scale
where, due to the gravitational corrections, the three gauge couplings
come together asymptotically free. We will determine the numerical
solution from (\ref{eq:RG equations inizio}) to (\ref{eq:RG equation fine})
for a range of values of $\rho,$ $\Lambda$ and $M_{R}$. The initial
conditions of the running parameters at the scale $\Lambda$ are given
by (\ref{eq:lambda h unificazione}) plus that for $y_{top}$ and
$y_{\nu}:$
\begin{equation}
y_{top}(\Lambda)=\frac{2}{\sqrt{3+\rho^{2}}}g_{2}(\Lambda),\,\,\,\,\, y_{\nu}(\Lambda)=\frac{2\rho}{\sqrt{3+\rho^{2}}}g_{2}(\Lambda)\,.
\end{equation}
The effective mass of the light neutrino is determined by the effective
coupling $\kappa$ and we choose to evaluate this mass at the scale
$M_{Z}$. Moreover, the running mass of the top quark to the ordinary
energies is given by 
\begin{equation}
M_{top}=\frac{1}{\sqrt{2}}y_{top}v_{0}
\end{equation}
 where $v_{0}\simeq246\, GeV$ is the vacuum expectation value of
the Higgs field. 

For the Higgs mass, we have to use the new relation due to the presence
of the new scalar field~\cite[Eq.35]{ccforgotafield},
\begin{equation}
M_{H}(M_{H})=v_{0}\sqrt{2\lambda_{H}(M_{H})\left(1-\frac{\lambda_{H\sigma}^{2}(M_{H})}{\lambda_{H}(M_{H})\lambda_{\sigma}(M_{H})}\right)}
\end{equation}
while the scalar-singlet $\sigma$ mass is proportional to its vacuum
expectation value $w_{0}$, near the Planck scale according to us,
through ~\cite[Eq.34]{ccforgotafield}, $M_{\sigma}^{2}=2\lambda_{\sigma}w_{0}^{2}+2v_{0}^{2}\lambda_{H\sigma}^{2}/\lambda_{\sigma}.$

\begin{figure}[H]
\begin{centering}
\includegraphics[width=16cm,height=5cm]{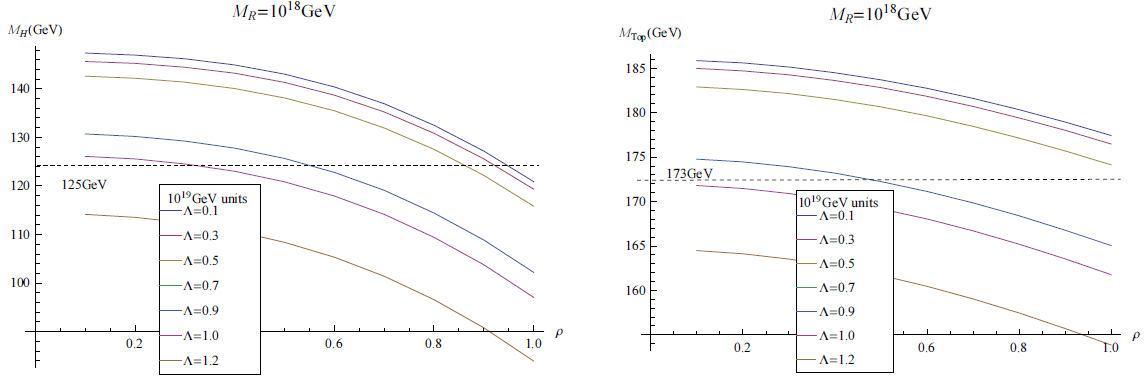}
\par\end{centering}

\centering{}\caption{{\small \label{fig:Mh ro lambda}}\textit{\small Higgs and top mass
in function of the parameter $\rho$ for seven different values of
$\Lambda.$ We can see that the Higgs mass around 125GeV and the top
mass around 173GeV constrain $\Lambda$ not over $1.0\,10^{19}GeV$. }}
\end{figure}

\begin{figure}[h]
\centering{}\includegraphics[width=16cm,height=5cm]{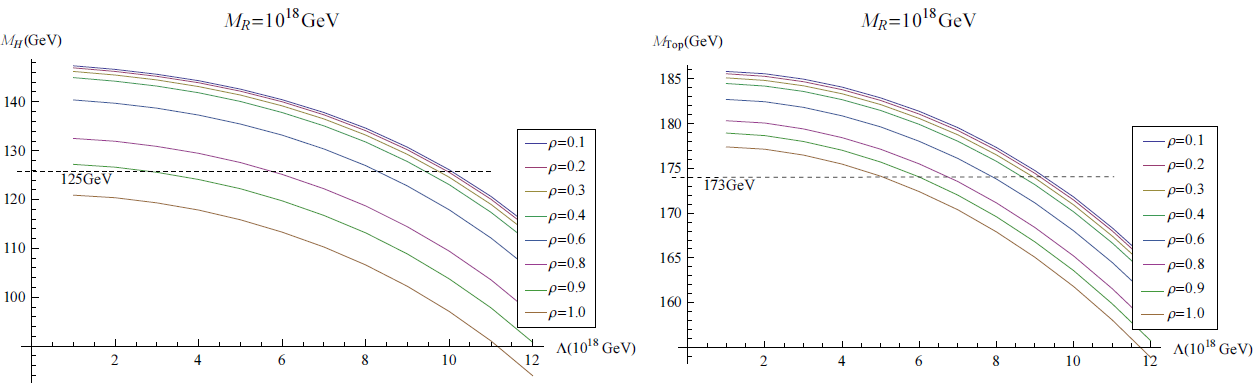}\caption{{\small \label{fig:mh lambda ro}}\textit{\small Higgs and top mass,
changing the unification parameter $\Lambda$ for eight different
values of $\rho$. Also in this case we can see that the Higgs mass
around 125GeV suggests an appropriate choice of $\rho$ not over $1.0$
meanwhile the top mass does not impose any constrain. Moreover both
$M_{H}$ and $M_{top}$ behaviors become $\rho$-indepent for $\rho\leq0.1$}}
\end{figure}

The results of the renormalization procedure for the Higgs and top
mass in terms of the three parameters $\rho,\,\Lambda,\, M_{R}$ are
shown in fig. \ref{fig:Mh ro lambda} and \ref{fig:mh lambda ro}.
In fig.~\ref{fig:Mh ro lambda} we see the Higgs and top mass values
in terms of $\rho$ for seven different values of $\Lambda$ and $M_{R}$
fixed: the Higgs mass around 125GeV and the top mass around 173GeV
suggest a consistent choice of $\Lambda$ not over $1.0\,10^{19}GeV$.
In fig. \ref{fig:mh lambda ro} is shown the behavior of the two masses
in function of $\Lambda$ for eight different values of $\rho$ with
$M_{R}$ fixed: also in this case we can see that the Higgs mass around
125GeV suggests an appropriate choice of $\rho$ not over $1.0$ meanwhile
the top mass does not impose any constrain. Moreover both $M_{H}$
and $M_{top}$ behaviors become $\rho$-indepent for $\rho\leq0.1$.
Moreover it is possible to verify that the parameter $M_{R}$ is not
important for the mass prediction since $M_{H}$ and $M_{top}$ grow
very slowly for its changes.  Therefore, in the end, we have a sensible
reduction on the choice of the three parameters values. 

\section{Conclusions}

In~\cite{DLM} the new singlet-scalar field $\sigma$, responsible
for the stability of the Higgs boson, has been derived spontaneously
from an high symmetry breaking that occurs at the Planck scale (that
means $w_{0}\simeq M_{P}$), mixing space-time spin and gauge degrees
of freedom. In the present work we have checked the possibility to
extend the unification scale up to the Planck scale with the presence
of the new scalar field non-minimal coupled to gravity.

We have, then, deduced a restriction of the free parameters of the
theory compatible with the Higgs and top mass: in particular we have
to take the parameters $\rho<1$ and $\Lambda$ not over $10^{19}\mbox{GeV}$.
However this constrain leaves some open problems: for $\Lambda\lesssim10^{19}GeV$
the three coupling constants are not exactly the same, although very
close: e.g. for $\Lambda=10^{19}\mbox{GeV}$ we have $g_{1}(\Lambda)^{2}=0.138$,
$g_{3}(\Lambda)^{2}=0.148$ and $g_{2}(\Lambda)^{2}=0.156$. Actually
we shall take at least $\Lambda\gtrsim3.0\times10^{19}\mbox{GeV}$
to have $g_{2}(\Lambda)^{2}=g_{3}(\Lambda)^{2}=g_{1}(\Lambda)^{2}=0.003$
and then to use consistently the spectral action at the fixed unification
scale. 

\begin{figure}[t]
\centering{}\includegraphics[width=9cm,height=5cm]{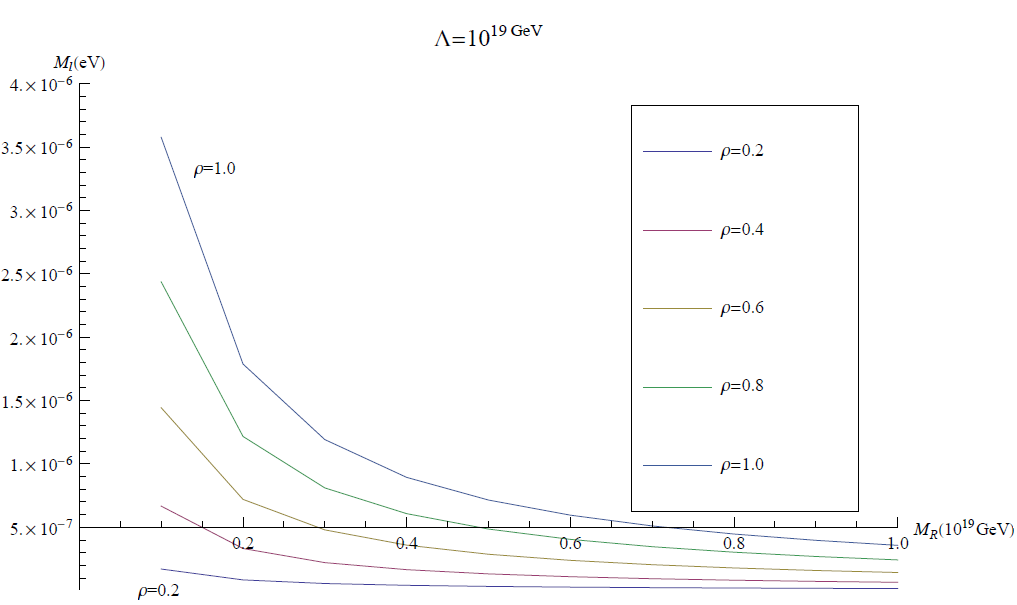}\caption{\label{fig:neutrino mass}\textit{\small Neutrino light mass, changing
the Majorana right mass value in the range $10^{18}\mbox{Gev}\,-\,10^{19}\mbox{Gev}$,
for five different values of $\rho$ and the unification scale $\Lambda$
fixed. We can see that the neutrino mass has a very low value, of
the order of $\mu eV$. Its value increases for increasing $\rho$
and for decreasing $M_{R}.$}}
\end{figure}
Moreover we have a neutrino mass problem which now becomes too small
since its light mass $m_{l}=\frac{1}{4}\kappa v_{0}^{2}$ is influenced
by $M_{R}$ in the denominator of $\kappa$ as in (\ref{eq: constant k});
as shown in fig. (\ref{fig:neutrino mass}) for $M_{R}\simeq10^{18}\mbox{GeV}$
the neutrino mass has a very low value of the order of $\mu eV$.
In order to rise the neutrino mass to few electronvolt, just two actions
are possible: (1) to increase the $\rho$ value, but nevertheless
it has an upper limit imposed by the Higgs and top mass; (2) to lower
the value of the Majorana right mass $M_{R}$ to $10^{14}GeV$. This
second possibility seems to indicate that the Majorana right mass
(proportional to the $\sigma$ v.e.v. $w_{0}$) responsible for the
seesaw mechanism, can not live at too high energy scales. This observation
suggests that we can not naively identify the scalar field $\sigma$
of the grand symmetry breaking~\cite{DLM} with the field that gives
mass to the Majorana right neutrino; otherwise, there may be some
mechanism that contributes to lower its mass, as in the case of neutrinos.
Beyond all, a more punctual analysis is required to investigate the
phenomenological consequences of this new and fascinating picture.

~

\textbf{Acknowledgments.} I would like to thank F. Lizzi for discussions
and suggestions.


\begin{thebibliography}{99}

\bibitem{Connesbook} A. Connes, \textit{Noncommutative Geometry}, Academic Press, 1984.      

\bibitem{Landi} G. Landi, {\it  An Introduction to Noncommutative Spaces and their Geometries}, {\sl Springer Lecture Notes in Physics 51}, Springer Verlag (Berlin Heidelberg) 1997. arXiv:hep-th/9701078.    

\bibitem{Ticos}  J.M.~Gracia-Bondia, J.C.~Varilly, H.~Figueroa, {\it     Elements of Noncommutative Geometry}, Birkhauser, 2000.

\bibitem{ConnesMarcolli} A. Connes, M. Marcolli, ``Noncommutative Geometry, Quantum Fields and Motives'', AMS 2007; 

\bibitem{ConnesLott}  A.~Connes and J.~Lott,   ``Particle Models And Noncommutative Geometry (expanded Version)''   Nucl.\ Phys.\ Proc.\ Suppl.\  {\bf 18B} (1991) 29.   

\bibitem{Schucker}  T.~Schucker,   ``Forces from Connes' geometry,''   Lect.\ Notes Phys.\  {\bf 659} (2005) 285   [hep-th/0111236].   

\bibitem{AC2M2} A.~H.~Chamseddine, A.~Connes and M.~Marcolli,   ``Gravity and the standard model with neutrino mixing''   Adv.\ Theor.\ Math.\ Phys.\  {\bf 11} (2007) 991   [arXiv:hep-th/0610241].   

\bibitem{Walterreview}    K.~van den Dungen and W.~D.~van Suijlekom,   ``Particle Physics from Almost Commutative Spacetimes''   arXiv:1204.0328 [hep-th].   

\bibitem{Jureit}   J.~H.~Jureit, T.~Krajewski, T.~Schucker and C.~A.~Stephan,   ``Seesaw and noncommutative geometry''   Phys.\ Lett.\ B {\bf 654} (2007) 127   [arXiv:0801.3731 [hep-th]].   

\bibitem{spectralaction} A.~H.~Chamseddine and A.~Connes,   ``The spectral action principle,''   Commun.\ Math.\ Phys.\  {\bf 186}, 731 (1997)   [arXiv:hep-th/9606001].   

\bibitem{ccforgotafield}  A.~H.~Chamseddine and A.~Connes,   ``Resilience of the Spectral Standard Model''   arXiv:1208.1030 [hep-ph].   

\bibitem{DLM}   A.~Devastato, F.~Lizzi and P.~Martinetti,   ``Grand Symmetry, Spectral Action, and the Higgs mass,''   arXiv:1304.0415 [hep-th].   



\bibitem{RobinsonWilczek}   S.~P.~Robinson and F.~Wilczek,   ``Gravitational correction to running of gauge couplings''   Phys.\ Rev.\ Lett.\  {\bf 96} (2006) 231601   [hep-th/0509050].   

\bibitem{Shaposhnikov:2009pv}   M.~Shaposhnikov and C.~Wetterich,   ``Asymptotic safety of gravity and the Higgs boson mass,''   Phys.\ Lett.\ B {\bf 683} (2010) 196   [arXiv:0912.0208 [hep-th]].   





\bibitem{Estrada:2012te}   C.~Estrada and M.~Marcolli,   ``Asymptotic safety, hypergeometric functions, and the Higgs mass in spectral action models''   arXiv:1208.5023 [hep-th].   

\bibitem{AndrianovLizzi}A.~A.~Andrianov and F.~Lizzi,   ``Bosonic Spectral Action Induced from Anomaly Cancelation,''   JHEP {\bf 1005} (2010) 057   [arXiv:1001.2036 [hep-th]].   

\bibitem{AndrianovKurkovLizzi}  A.~A.~Andrianov, M.~A.~Kurkov and F.~Lizzi,   ``Spectral action, Weyl anomaly and the Higgs-Dilaton potential,''   JHEP {\bf 1110} (2011) 001   [arXiv:1106.3263 [hep-th]].   

\bibitem{Kurkov:2012dn} M.~A.~Kurkov and F.~Lizzi,   ``Higgs-Dilaton Lagrangian from Spectral Regularization,''   Mod.\ Phys.\ Lett.\ A {\bf 27} (2012) 1250203   [arXiv:1210.2663 [hep-th]].   


\bibitem{Gilkey} P. Gilkey, \textit{Invariance Theory, the Heat     Equation and the Athiya-Singer Index Theorem}, Publish or     Perish, 1984.

\bibitem{Vassilevich} D.~V.~Vassilevich,   ``Heat kernel expansion: User's manual''   Phys.\ Rept.\  {\bf 388} (2003) 279   [hep-th/0306138].   


\bibitem{CCpart1}A.~H.~Chamseddine and A.~Connes,   ``Noncommutative Geometry as a Framework for Unification of all Fundamental Interactions including Gravity. Part I''   Fortsch.\ Phys.\  {\bf 58} (2010) 553   [arXiv:1004.0464 [hep-th]].   

\bibitem{Marcolli:2009in}   M.~Marcolli and E.~Pierpaoli,   ``Early Universe models from Noncommutative Geometry,''   Adv.\ Theor.\ Math.\ Phys.\  {\bf 14} (2010)   [arXiv:0908.3683 [hep-th]].   

\bibitem{AndrianovEspriu}   A.~A.~Andrianov, D.~Espriu, M.~A.~Kurkov and F.~Lizzi,   ``Universal Landau Pole''   arXiv:1302.4321 [hep-th].   

\bibitem{Maiani}   L.~Maiani, G.~Parisi and R.~Petronzio,   ``Bounds on the Number and Masses of Quarks and Leptons''   Nucl.\ Phys.\ B {\bf 136} (1978) 115.   



\bibitem{AppelquistCarazzone}   T.~Appelquist and J.~Carazzone,   ``Infrared Singularities and Massive Fields''   Phys.\ Rev.\ D {\bf 11} (1975) 2856.   

\bibitem{Antusch}   S.~Antusch, J.~Kersten, M.~Lindner and M.~Ratz,   ``Neutrino mass matrix running for nondegenerate seesaw scales,''   Phys.\ Lett.\ B {\bf 538} (2002) 87   [hep-ph/0203233].   

\bibitem{Machacek2}   M.~E.~Machacek and M.~T.~Vaughn,   ``Two Loop Renormalization Group Equations in a General Quantum Field Theory. 2. Yukawa Couplings''   Nucl.\ Phys.\ B {\bf 236} (1984) 221.   

\bibitem{Machacek3}   M.~E.~Machacek and M.~T.~Vaughn,   ``Two Loop Renormalization Group Equations in a General Quantum Field Theory. 3. Scalar Quartic Couplings,''   Nucl.\ Phys.\ B {\bf 249} (1985) 70.   


\end{thebibliography}
\end{document}